\documentclass[fleqn,11pt,twoside]{article}

\usepackage{amsthm,amsthm,amssymb, color, xcolor,epsfig, graphics, subfigure}

\usepackage{amsmath, graphicx, latexsym, lscape}

\makeatletter
\newcommand{\copyrightnote}[2]{{\renewcommand{\thefootnote}{}
 \footnotetext{\small\it
\begin{flushleft}
 \copyright \ #1   #2  
\end{flushleft}}}}

\newcommand{\Name}[1]{\begin{flushleft}
                       \LARGE \bf #1
                       \end{flushleft}\vspace{-3mm}}

\newcommand{\Author}[1]{\begin{flushleft}
                       \it #1 \end{flushleft}}

\newcommand{\Address}[1]{\begin{flushleft}
                       \it #1 \end{flushleft}}

\newcommand{\Date}[1]{\begin{flushleft}
                      \small  \it #1 \end{flushleft}}

\newcommand{\evenhead}{Author \ name}
\newcommand{\oddhead}{Article \ name}

\renewcommand{\@evenhead}{
\hspace*{-3pt}\raisebox{-15pt}[\headheight][0pt]{\vbox{\hbox to \textwidth
{\thepage \hfil \evenhead}\vskip4pt \hrule}}}
\renewcommand{\@oddhead}{
\hspace*{-3pt}\raisebox{-15pt}[\headheight][0pt]{\vbox{\hbox to \textwidth
{\oddhead \hfil \thepage}\vskip4pt\hrule}}}
\renewcommand{\@evenfoot}{}
\renewcommand{\@oddfoot}{}

\long\def\@makecaption#1#2{%
  \vskip\abovecaptionskip
  \sbox\@tempboxa{\small \textbf{#1.}\ \ #2}%
  \ifdim \wd\@tempboxa >\hsize
    {\small \textbf{#1.}\ \ #2}\par
  \else
    \global \@minipagefalse
    \hb@xt@\hsize{\hfil\box\@tempboxa\hfil}%
  \fi
  \vskip\belowcaptionskip}

\newcommand{\JNMPnumberwithin}[3][\arabic]{%
  \@ifundefined{c@#2}{\@nocounterr{#2}}{%
    \@ifundefined{c@#3}{\@nocnterr{#3}}{%
      \@addtoreset{#2}{#3}%
      \@xp\xdef\csname the#2\endcsname{%
        \@xp\@nx\csname the#3\endcsname .\@nx#1{#2}}}}%
}

\newcommand{\resetfootnoterule} {
  \renewcommand\footnoterule{%
  \kern-3\p@
  \hrule\@width.4\columnwidth
  \kern2.6\p@}
}

\renewcommand{\footnoterule}{}

\makeatother

\theoremstyle{definition}

\newtheorem{prop}{Proposition}[section]

\begin{document}

\renewcommand{\evenhead}{ {\LARGE\textcolor{blue!10!black!40!green}{{\sf \ \ \ ]ocnmp[}}}\strut\hfill X-B Hu, G Yu and Y Zhang}
\renewcommand{\oddhead}{ {\LARGE\textcolor{blue!10!black!40!green}{{\sf ]ocnmp[}}}\ \ \ \ \  Integrable discretization, recursion operators and soliton hierarchies}

\thispagestyle{empty}
\newcommand{\FistPageHead}[3]{
\begin{flushleft}
\raisebox{8mm}[0pt][0pt]
{\footnotesize \sf
\parbox{150mm}{{Open Communications in Nonlinear Mathematical Physics}\ \ \ \ {\LARGE\textcolor{blue!10!black!40!green}{]ocnmp[}}
\quad Special Issue 1, 2024\ \  pp
#2\hfill {\sc #3}}}\vspace{-13mm}
\end{flushleft}}

\FistPageHead{1}{\pageref{firstpage}--\pageref{lastpage}}{ \ \ }

\strut\hfill

\strut\hfill

\copyrightnote{The author(s). Distributed under a Creative Commons Attribution 4.0 International License}

\begin{center}
{  {\bf This article is part of an OCNMP Special Issue\\ 
\smallskip
in Memory of Professor Decio Levi}}
\end{center}

\smallskip

\Name{Integrable discretization of recursion operators and unified bilinear forms to soliton hierarchies}

\Author{Xingbiao Hu$^{\,12}$,  Guofu Yu$^{\,3}$ and Yingnan Zhang$^{\,4}$}

\Address{$^{1}$ LSEC, Institute of Computational Mathematics and Scientific Engineering Computing,Academy of Mathematics and Systems Science, Chinese Academy of Sciences,  Beijing, China\\[2mm]
$^{2}$ School  of Mathematical Sciences, University of the Chinese Academy of Sciences, Beijing, China\\[2mm]
$^{3}$ Department of Mathematics, Shanghai Jiao Tong University, Shanghai, China\\[2mm]
$^{4}$Key Laboratory for NSLSCS, Ministry of Education, School of Mathematical Sciences, Nanjing Normal University, Nanjing, Jiangsu, China\\[2mm]
}

\Date{Received August 22, 2023; Accepted December 24, 2023}

\setcounter{equation}{0}

\begin{abstract}

\noindent 
In this paper, we give a procedure  for discretizing recursion operators by utilizing unified bilinear forms within integrable hierarchies. To illustrate this approach, we present unified bilinear forms for both the AKNS hierarchy and the KdV hierarchy, derived from their respective recursion operators. Leveraging the inherent connection between soliton equations and their auto-B\"acklund transformations, we discretize the bilinear integrable hierarchies and derive discrete recursion operators. These discrete recursion operators exhibit convergence towards the original continuous forms when subjected to a standard limiting process.

\end{abstract}

\label{firstpage}


\section{Introduction}

The main objective of this paper is to address the challenge of discretizing the recursion operator of a given continuous integrable equation while simultaneously ensuring that the discretized equation maintains its integrable nature. This entails transforming the continuous recursion operator into a discrete version that serves as the recursion operator for the discrete equation. To achieve this goal, we will begin by revisiting pertinent background information and contextual elements related to this subject.

Although not rigorously proven, it is widely accepted that a fully integrable evolutionary equation is inherently part of a set of equations that commute, forming a family or hierarchy. Each member within this hierarchy is also fully integrable. The equations within a hierarchy exhibit several shared integrable characteristics. For instance, they possess a common spectral problem, and their solutions share analogous structures. From a computational perspective, simplifying the representation of the entire integrable hierarchy is necessary, as higher-order members of the hierarchy often become unwieldy in their expression.

Within the existing literature, there are at least two methods available for achieving this simplification, thereby facilitating the study of the properties inherent in the complete integrable hierarchy.

One approach involves expressing the equations within a hierarchy in a consolidated bilinear form utilizing Hirota's bilinear derivatives \cite{hirota71}. Hirota's bilinear form was initially introduced for constructing multi-soliton solutions \cite{hirota71,hirota721,hirota722,hirota73}. This technique, known as Hirota's direct method, proves highly effective for deriving soliton solutions. By employing an appropriate transformation of dependent variables and introducing an infinite set of variables ($x=t_1, t_2, t_3, \ldots$), equations from the same hierarchy can be reformulated into a unified bilinear form.
It's important to acknowledge that the unified bilinear form for the AKNS hierarchy was initially derived by Newell \cite{Newell}. During the 1980s, the Sato school made significant strides by establishing bilinear KP and BKP hierarchies, alongside their corresponding solutions through the $\tau$ function, using the Kac-Moody algebraic representation \cite{SA,DKJM,JM}. This breakthrough marked substantial progress. Additionally, bilinear equations for the KdV and mKdV hierarchies were obtained through elementary methods in \cite{Ma1,Ma2,Ma3,Ma4}. The unified explicit expression holds significant advantages. For instance, in \cite{Hu1}, the corresponding B\"acklund transformations and nonlinear superposition methods for the KdV and mKdV hierarchies were derived from the concise unified expression. Similarly, in \cite{Hu2}, rational solutions for the classical Boussinesq hierarchy were obtained based on its unified bilinear form.

Another approach involves expressing the integrable equation family in a recursive manner using a concept known as a recursion operator \cite{olver,zk}. Recursion operators were first introduced by Olver in 1977 \cite{olver}, and their development continued through the works of Fuchssteiner \cite{Fuchssteiner} and Fokas and Santini \cite{fs,sf}.
The recursive structure is a sophisticated attribute of completely integrable systems, with the recursion operator, as mentioned in \cite{zk}, assuming a pivotal role in defining these recursive properties. This operator enables the concise formulation of the integrable equation family and determines the collection of Hamiltonian structures associated with it. In essence, a recursion operator serves as the generative operator for the integrable equation family, concurrently playing the same role for the Hamiltonian structure family.
For further insights, one can refer to \cite{zk,olver,akns,Fuchssteiner,fs,sf,fn,ff,fa,fg,oevel,wang1,wang2,wang2019}, along with their respective references, to delve into the details of this concept and its various applications.

When presented with an integrable hierarchy alongside its associated recursion operator, it is possible to derive the unified bilinear form by employing a suitable variable transformation. In \cite{Hu1}, for instance, the unified bilinear forms of the KdV hierarchy, mKdV hierarchy, and classical Boussinesq hierarchy were obtained using their respective recursion operators. Utilizing the recursion operator to deduce the bilinear form offers numerous advantages, such as reduced computational complexity and a more straightforward unified and explicit expression.

Conversely, when the unified bilinear form is known, it is also possible to determine the recursion operator using the same type of transformation, as discussed in \cite{Hu3,Hu4}. This  relationship between the recursion operator and the unified bilinear form contributes to a comprehensive understanding of the integrable hierarchy's properties and behaviors.


We now return to the issue of integrable discretization.
Traditionally, the challenge of integrable discretization revolves around transforming an integrable system into a discrete counterpart while preserving its integrability. Pioneering efforts in this area were made more than 35 years ago by researchers such as Ablowitz and Ladik \cite{AL}, Hirota \cite{hr1,hr2,hr3}, Nijhoff \cite{ncwq,qncl}, and Levi \cite{levi1,levi2}. These individuals laid the foundation for exploring this issue.
Additional contributions have come from various approaches, such as Suris' Hamiltonian approach \cite{sur} and Schiff's loop group approach \cite{schiff}, among others. The references provided are just a few representative examples, as numerous other recent works delve into this topic as well. For deeper investigation, additional progress and comprehensive research studies can be found in publications like \cite{bs,ablowitz2004, Hietarinta2016, levi2023, feng1, feng2, suris1, Tsuchida, veni, yu1, yu2, zhang1, zhang2, zhang3, zhang4, liu1, zhu1}, along with their respective reference sections. These works collectively contribute to the evolving understanding of integrable discretization methodologies.

In their book \cite{bs}, Bobenko and Suris introduced a broader perspective on integrable discretization, emphasizing the importance of "discretizing the whole theory, not just the equations."  Motivated by this insightful viewpoint, it becomes both logical and meaningful to extend integrable discretization to include recursion operators. This paper aims to demonstrate a systematic procedure for achieving this.
To illustrate this approach, we will employ two prominent examples: the AKNS hierarchy and the KdV hierarchy. Our methodology involves discretizing not only the unified bilinear forms of these hierarchies but also their recursion operators. The central technique for discretization is rooted in the well-established concept that integrable discretizations can be realized through a suitable interpretation of B\"acklund transformations \cite{levi1,levi2}.

In the forthcoming sections, we will delve into the specific details of discretizing the AKNS hierarchy and the KdV hierarchy, elaborating on the process and showcasing the outcomes of this integrable discretization approach.

\section{Recursion operator and unified bilinear form}
\subsection{AKNS hierarchy}
\noindent From the spectral problem
\begin{eqnarray}
\Psi_x=\left( \begin{array}{ll} -i\lambda & q(x,t)\\ r(x,t) &\quad
i\lambda
\end{array}\right)\Psi,\qquad \Psi=\left( \begin{array}{l} \psi_1
\\\psi_2 \end{array}\right) \label{sp}
\end{eqnarray}
we can obtain the following AKNS hierarchy
\cite{akns,calogero, Ablow2,Newell}
\begin{eqnarray}
\left(\begin{array}{l} q_t \\ r_t\end{array}\right)=L^m
\left(\begin{array}{l} q_x \\ r_x\end{array}\right),\qquad (m\geq 1)
\end{eqnarray}
where
\begin{eqnarray}
L=\frac{1}{2i}\left( \begin{array}{ll} -\partial_x+2q\partial_x^{-1}r
& \quad 2q\partial_x^{-1}q\\
-2r\partial_x^{-1}r &\partial_x-2r\partial_x^{-1}q
\end{array}\right).  \label{operator_akns}
\end{eqnarray}
Introducing infinite time variables $x=t_1, t_2, t_3,\cdots$ and
regard $q, r$ as functions of $t=t(t_1,t_2,t_3\cdots)$, we have the
following equivalent equation
\begin{eqnarray}
\left(\begin{array}{l} q_{t_{m+1}} \\ r_{t_{m+1}}\end{array}\right)
=\frac{1}{2i}\left( \begin{array}{ll} -\partial_x+2q\partial_x^{-1}r
& \quad 2q\partial_x^{-1}q\\
-2r\partial_x^{-1}r &\partial_x-2r\partial_x^{-1}q
\end{array}\right)\left(\begin{array}{l} q_{t_{m}} \\
r_{t_{m}}\end{array}\right). (m \geq 1)\label{akns-c}
\end{eqnarray}
or equivalently
\begin{eqnarray}
&& q_{t_{m+1}}=-\frac{1}{2i}q_{xt_m}-iq\partial_x^{-1}(qr)_{t_m},\label{akns-non1}\\
&& r_{t_{m+1}}=\frac{1}{2i}r_{xt_m}+ir\partial_x^{-1}(q
r)_{t_m}.\label{akns-non2}
\end{eqnarray}
By variable transformation
\begin{eqnarray}
q=\frac{\sigma}{\tau},\qquad r=\frac{\rho}{\tau}
\end{eqnarray}
nonlinear equations (\ref{akns-non1})-(\ref{akns-non2}) can be changed into
\begin{eqnarray}
&& (D_{t_{m+1}}-\frac{i}{2}D_{t_1}D_{t_m})\sigma\cdot \tau=0,\label{akns-b1}\\
&& (D_{t_{m+1}}+\frac{i}{2}D_{t_1}D_{t_m})\rho\cdot \tau=0,\label{akns-b2}\\
&& D_{t_1}^2\tau \cdot \tau=-2\sigma\rho.\label{akns-b3}
\end{eqnarray}
Here the $D$-operator is defined by
\begin{eqnarray}
&& D_t^mD_x^n a(t,x)\cdot b(t,x)=\frac{\partial^m}{\partial s^m}\frac{\partial^n}{\partial y^n}a(t+s,x+y)b(t-s,x-y)|_{s=0,y=0},\nonumber\\
&& \qquad \qquad \qquad \qquad \qquad \qquad \qquad \qquad \qquad
m,n=0,1,2,\cdots,
\end{eqnarray}
or by the exponential identity
\begin{eqnarray}
&& exp( \delta D_z)a(z)\cdot b(z)=exp( \delta \partial_y)(a(z+y)b(z-y))|_{y=0},\nonumber\\
&& \qquad \qquad \qquad \qquad =a(z+\delta)b(z-\delta).
\end{eqnarray}
In fact, (\ref{akns-non1}) can be written as
\begin{eqnarray}
&&\frac{D_{t_{m+1}}\sigma\cdot \tau}{\tau^2}=-\frac{1}{2i}
\frac{(\sigma_x\tau-\sigma\tau_x)_{t_m}\tau^2-2\tau\tau_{t_m}
(\sigma_x\tau-\sigma\tau_x)}{\tau^4}-i\frac{\sigma}{\tau}\partial_x^{-1}
\left(\frac{\sigma\rho}{\tau^2}\right)_{t_n}\nonumber\\
&&=-\frac{1}{2i}\frac{(\sigma_{xt_m}\tau+\sigma_x\tau_{t_m}
-\sigma_{t_m}\tau_x-\sigma\tau_{xt_m})\tau^2-2\tau\tau_{t_m}(\sigma_x\tau-\sigma\tau_x)}
{\tau^4}\\
&&\qquad+\frac{i\sigma}{2\tau}\partial_x^{-1}\left(
\frac{D_{t_1}^2\tau\cdot \tau}{\tau^2}\right)_{t_m}\nonumber\\
&&=-\frac{1}{2i}\frac{\sigma_{xt_m}\tau+\sigma_x\tau_{t_m}
-\sigma_{t_m}\tau_x-\sigma\tau_{xt_m}}{\tau^2}\nonumber\\
&&=-\frac{1}{2i}\frac{D_xD_{t_m}\sigma\cdot \tau}{\tau^2},
\end{eqnarray}
so (\ref{akns-b1}) is obtained. We can deduce (\ref{akns-b2}) from
(\ref{akns-non2}) similarly.  We should remark here that the bilinear
form of AKNS hierarchy has been given in \cite{Newell}, but there is no
recursion operator used explicitly.

\begin{prop}
A bilinear B\"acklund transformation for AKNS hierarchy (\ref{akns-b1})-(\ref{akns-b3})is
\begin{eqnarray}
&&D_x\tau\cdot\hat{\tau}=\lambda \sigma\hat{\rho}+\mu \tau\hat{\tau},\label{akns-bt1}\\
&&\lambda D_x\sigma\cdot\hat{\tau}+\gamma \sigma\hat{\tau}-\tau\hat{\sigma}=0,\label{akns-bt2}\\
&&\lambda D_x\tau\cdot\hat{\rho}+\gamma \tau\hat{\rho}-\rho\hat{\tau}=0,\label{akns-bt3}\\
&&(2iD_{t_{m+1}}-\frac{1}{2}D_xD_{t_n}-(\lambda^{-1}\gamma+\frac{1}{2}\mu )D_{t_m})\tau\cdot\hat{\tau}=\frac{1}{2}\lambda D_{t_m}\sigma\cdot\hat{\rho}.\label{akns-bt4}
\end{eqnarray}
where $\sigma$, $\rho$, $\tau$ and $\hat{\sigma}$, $\hat{\rho}$, $\hat{\tau}$ are two sets of solutions of the AKNS hierarchy
and $\lambda$, $\gamma$, $\mu$ are B\"acklund parameters.
\end{prop}

\subsection{KdV hierarchy}
The bilinear KdV hierarchy is
\begin{eqnarray}
&& (4D_xD_{t_3}+D_x^4)f\cdot f=0,\label{bkdv1}\\
&& (D_xD_{t_{2m+1}}+\frac{1}{6}D_{t_{2m-1}}D_x^3-\frac{1}{3}D_{t_3}D_{t_{2m-1}})f\cdot f=0,\label{bkdv-h}
\end{eqnarray}
where $ m=1,2,\cdots,t_1=x$.
When $m=1$, (\ref{bkdv-h}) reduces to (\ref{bkdv1}) which is just the general KdV equation.
With $u=2(\ln f)_{xx}$, (\ref{bkdv1}) and (\ref{bkdv-h}) give
\begin{eqnarray}
&& 4u_{t_3}+u_{xxx}+6uu_x=0,\label{kdv1}\\
&& 4u_{t_{2m+1}}+Lu_{t_{2m-1}}=0.\label{smooth operator1}
\end{eqnarray}
where $L=\partial_x^2+4u+2u_x\partial_x^{-1}$ is the recursion operator.

\begin{prop}
A bilinear B\"acklund transformation for KdV hierarchy (\ref{kdv1})-(\ref{smooth operator1})is
\begin{eqnarray}
&&(D_x^2-\lambda D_x)f\cdot g=0,\label{kdv-bt1}\\
&&(4D_{t_3}+D_x^3)f\cdot g=0,\label{kdv-bt2}\\
&&(D_{t_{2m+1}}+\frac{1}{4}D_{t_{2m-1}}D_x^2-\frac{\lambda}{4}D_xD_{t_{2m-1}}+\frac{\lambda^2}{4}D_{t_{2m-1}})f\cdot g=0,\label{kdv-bt3}
\end{eqnarray}
where $f$, $g$ are two solutions of KdV hierarchy
and $\lambda$ is B\"acklund parameter.
\end{prop}

\section{Integrable discretization}
There are many methods to derive integrable analogues of soliton equations \cite{AL,AL1,hr1,hr2,hr3,ncwq,qncl,levi1,levi2,sur}. In this paper we discretize the integrable hierarchies based on the compatibility between an integrable system and its auto-B\"acklund transformation (BT).

\subsection{Discrete AKNS hierarchy}
The study of discrete AKNS system started in 1970s, when the inverse scattering method \cite{akns,Ablow2,calogero} was used to solve a number of
difference evolution equations\cite{AL1,akns51,akns52,akns8,akns9,akns11}.  There are two different discrete analogues of the AKNS equation. One is given by Ablowitz and Ladik through a discretized version of the eigenvalue problem of Zakharov and Shabat.  There are many applications of this discrete system, especially in the discrete NLS equation which is known as the Ablowitz-Ladik hierarchy. Recent study about the discrete AKNS system can be found in \cite{zhangdajun1}.

The other  was given by   Chudnovsky  through B\"acklund transformation of the AKNS system \cite{akns12, akns13}.  The multi-component discrete AKNS can be found in \cite{akns14}. The recursion operator was proposed in \cite{akns15,akns16} through mastersymmetry approach and Lax pair analysis. In the present paper, we study the same discrete system as that in \cite{akns13,akns15} by discretizing the unified bilinear form. The discrete recursion operator can be obtained directly from the discrete unified bilinear form.

The discretization of the AKNS hierarchy (\ref{akns-b1})-(\ref{akns-b3}) can be written in a unified bilinear form
\begin{eqnarray}
&&D_x\tau_n\cdot\tau_{n-1}=-h \sigma_n\rho_{n-1},\label{dakns1}\\
&&D_x\sigma_n\cdot\tau_{n-1}=\frac{\sigma_n\tau_{n-1}-\tau_n\sigma_{n-1}}{h},\label{dakns2}\\
&&D_x\tau_n\cdot\rho_{n-1}=\frac{\tau_n\rho_{n-1}-\rho_n\tau_{n-1}}{h},\label{dakns3}\\
&&(iD_{t_2}+\frac{1}{2}D_x^2)\sigma_n\cdot \tau_n=0,\label{dakns4}\\
&&(iD_{t_2}-\frac{1}{2}D_x^2)\rho_n\cdot \tau_n=0,\label{dakns5}\\
&&D_x^2\tau\cdot \tau=-2\sigma_n\rho_n,\label{dakns6}\\
&&(D_{t_{m+1}}-\frac{i}{2}D_{x}D_{t_m})\sigma_n\cdot \tau_n=0,\label{dakns7}\\
&&(D_{t_{m+1}}+\frac{i}{2}D_{x}D_{t_m})\rho_n\cdot \tau_n=0,\label{dakns8}
\end{eqnarray}
where (\ref{dakns1})-(\ref{dakns3}) come from the B\"acklund transformation (\ref{akns-bt1})-(\ref{akns-bt3}) with $\mu=0$, $\gamma=1$, $\lambda=-h$, $\tau=\tau_{n}$, $\sigma=\sigma_{n}$, $\rho=\rho_{n}$, $\hat{\tau}=\tau_{n-1}$, $\hat{\sigma}=\sigma_{n-1}$, $\hat{\rho}=\rho_{n-1}$, while (\ref{dakns4})-(\ref{dakns8}) come from the original AKNS hierarchy (\ref{akns-b1})-(\ref{akns-b3}).

With $v_n=(\ln \tau_n)_x$, $q_n=\frac{\sigma_n}{\tau_n}$, $r_n=\frac{\rho_n}{\tau_n}$, the above system transforms into
\begin{eqnarray}
&&v_n-v_{n-1}=-hq_nr_{n-1},\\
&&q_{n,x}=\frac{q_n-q_{n-1}}{h}-q_n(v_n-v_{n-1})=\frac{q_n-q_{n-1}}{h}+hq_n^2r_{n-1},\\
&&r_{n-1,x}=\frac{r_n-r_{n-1}}{h}-hq_nr_{n-1}^2,\\
&&q_{n,t_2}-\frac{i}{2}q_{n,xx}+iq_n^2r_n=0,\\
&&r_{n,t_2}+\frac{i}{2}r_{n,xx}-iq_nr_n^2=0,\\
&&q_{n,t_{m+1}}-\frac{i}{2}q_{n,xt_m}-iq_nv_{n,t_m}=0,\\
&&r_{n,t_{m+1}}+\frac{i}{2}r_{n,xt_m}+ir_nv_{n,t_m}=0.
\end{eqnarray}
Eliminating $q_{n,x}$, $q_{n,xx}$, $r_{n,x}$, $r_{n,xx}$ and $v_{n,t_m}$, we have
\begin{eqnarray}
&&q_{n,t_2}+iq_n^2r_n-\frac{i}{2}(\frac{q_n-2q_{n-1}+q_{n-2}}{h^2}+2q_n^2r_{n-1}-2q_nq_{n-1}r_{n-1}\nonumber\\
&&\qquad\qquad\qquad+q_n^2r_n-q_{n-1}^2r_{n-2}+h^2q_n^3r_{n-1}^2)=0,\label{dakns9}\\
&&r_{n,t_2}-iq_nr_n^2+\frac{i}{2}(\frac{r_{n+2}-2r_{n+1}+r_{n}}{h^2}+2q_{n+1}r_{n}^2-2q_{n+1}r_{n+1}r_n\nonumber\\
&&\qquad\qquad\qquad-q_{n+2}r_{n+1}^2+q_nr_n^2+h^2q_{n+1}^2r_{n}^3)=0.\label{dakns10}
\end{eqnarray}
and
\begin{equation}
\left( \begin{array}{c}
q_n\\
r_n
\end{array} \right)_{t_{m+1}}=\left(
\begin{array}{cc}
A & B  \\
C & D
\end{array}\right)
\left( \begin{array}{c}
q_n\\
r_n
\end{array} \right)_{t_{m}},\quad m\geq 2\label{dakns11}
\end{equation}
where
\begin{eqnarray}
A&=&\frac{1}{2i}[-\frac{1}{h}(1-E^{-1})-2hq_nr_{n-1}+2hq_n(E-1)^{-1}r_nE]\\
B&=&\frac{1}{2i}[-hq_n^2E^{-1}+2hq_n(E-1)^{-1}q_{n+1}]\\
C&=&\frac{1}{2i}[-hr_n^2E-2hr_n(E-1)^{-1}r_{n}E]\\
D&=&\frac{1}{2i}[\frac{1}{h}(E-1)-2hq_{n+1}r_{n}-2hr_n(E-1)^{-1}q_{n+1}].
\end{eqnarray}
and $E$ is the shift operator defined by $E(f_n):=f_{n+1}$ for arbitrary functions of $n$.
As $h\rightarrow 0$, $\frac{1}{h}(1-E^{-1})\rightarrow \partial_x$, $\frac{1}{h}(E-1)\rightarrow \partial_x$ and $h(E-1)^{-1}\rightarrow \partial_x^{-1}$, thus
\begin{eqnarray}
A&\rightarrow&\frac{1}{2i}[-\partial_x+2q\partial_x^{-1}r]\\
B&\rightarrow&\frac{1}{2i}[2q\partial_x^{-1}q]\\
C&\rightarrow&\frac{1}{2i}[-2r\partial_x^{-1}r]\\
D&\rightarrow&\frac{1}{2i}[\partial_x-2r\partial_x^{-1}q],
\end{eqnarray}
which indicates the relation between the discrete hierarchy (\ref{dakns11}) and its corresponding continuous form(\ref{akns-c}). 
Therefore, from these discussions we can give the following proposition.
\begin{prop}
The differential-difference system (\ref{dakns9})-(\ref{dakns11}) is an intgerable discretization of the AKNS hierarchy (\ref{akns-c}). 
As $h\rightarrow 0$, the discrete  operator in (\ref{dakns11}) converges to the recursion operator (\ref{operator_akns}).
\end{prop}

However, it is essential to emphasize that the derived recursion operator is different with the operator previously documented in \cite{akns14,akns15,akns16}.
The discrete recursion operator in this study  primarily serves a formal recursive role. 
The determination of whether this operator truly exhibits a strong symmetry for the integrable discretizations of the AKNS hierarchy necessitates further verification.

\subsection{Discrete KdV hierarchy}
There are several different discrete analogues of the KdV equations. One is given by Nijhoff and Capel which reduces to the continues KdV equation  in a non-standard limit \cite{nijhoff}.  Another is given by Ablowitz and Ladik by discretizing the eigenvalue problem \cite{AL1} and by Hirota and Ohta  through discretizing the bilinear operators \cite{hr1,ohta}. This discrete KdV equation can also be interpreted as a generalized Volterra system \cite{sur,akns8}. In \cite{schiff}, Schiff give a full discretized KdV hierarchy by loop group approach.  Different from the discrete version given by Nijhoff, the last two discrete KdV equations  reduce to the continues KdV equation in a standard limit.  In the present paper, we give a different discrete analogue of the KdV hierarchy by discretizing the unified bilinear form and derive a discrete recursion operator directly from the discrete unified bilinear form. In a standard limit, the discrete hierarchy and the discrete recursion operator reduce to the continues KdV hierarchy and resursion operator.

The discretization of the KdV hierarchy (\ref{bkdv1})-(\ref{bkdv-h}) can be written in a unified bilinear form
\begin{eqnarray}
&& (D_x^2-\frac{2}{h}D_x)f_{n+1}\cdot f_n=0,\label{db1}\\
&& (4D_xD_{t_3}+D_x^4)f_n\cdot f_n=0,\label{db2}\\
&& (D_xD_{t_{2m+1}}+\frac{1}{6}D_{t_{2m-1}}D_x^3-\frac{1}{3}D_{t_3}D_{t_{2m-1}})f_n\cdot f_n=0.\label{db3}
\end{eqnarray}
Setting $w_n=2\log f_n$, $v_n=w_{n,x}$, $u_n=v_{n,x}$, $p_n=u_{n,x}$, $q_n=p_{n,x}$, $r_n=q_{n,x}$, (\ref{db1})-(\ref{db3}) transform into
\begin{eqnarray}
&& \delta_{+}u_n=\frac{2}{h}\delta_{-}v_n-\frac{1}{2}(\delta_{-}v_n)^2,\label{d1}\\
&& \delta_{+}p_n=\frac{2}{h}\delta_{-}u_n-(\delta_{-}u_n)(\delta_{-}v_n),\label{d2}\\
&& \delta_{+}q_n=\frac{2}{h}\delta_{-}p_n-(\delta_{-}u_n)^2-(\delta_{-}p_n)(\delta_{-}v_n),\label{d3}\\
&& \delta_{+}r_n=\frac{2}{h}\delta_{-}q_n-3(\delta_{-}p_n)(\delta_{-}u_n)-(\delta_{-}q_n)(\delta_{-}v_n),\label{d4}\\
&& 4v_{n,t_3}+q_n+3u_n^2=0,\label{d5}\\
&& 4u_{n,t_3}+r_n+6p_nu_n=0,\label{d6}\\
&& v_{n,t_{2m+1}}+\frac{1}{6}(p_{n,t_{2m-1}}+3u_nv_{n,t_{2m-1}})-\frac{1}{3}w_{n,t_3,t_{2m-1}}=0,\label{d7}\\
&& u_{n,t_{2m+1}}+\frac{1}{6}(q_{n,t_{2m-1}}+3p_nv_{n,t_{2m-1}}+3u_nu_{n,t_{2m-1}})-\frac{1}{3}v_{n,t_3,t_{2m-1}}=0,\label{d8}
\end{eqnarray}
here $\delta_{+}f_n:=f_{n+1}+f_n$ and $\delta_{-}f_n:=f_{n+1}-f_n$ for arbitrary functions of $n$.

Eliminating $t_3$ from (\ref{d5}) and (\ref{d8}), we have
\begin{eqnarray*}
&& 4u_{n,t_{2m+1}}+q_{n,t_{2m-1}}+2p_nv_{n,t_{2m-1}}+4u_nu_{n,t_{2m-1}}=0,
\end{eqnarray*}

(\ref{d5}), (\ref{d7}) effected by $\delta_{-}$ gives
\begin{eqnarray}
&& 4\delta_{-}v_{n,t_3}+\delta_{-}q_n+3(\delta_{+}u_n)(\delta_{-}u_n)=0,\label{1}\\
&& \delta_{-}v_{n,t_{2m+1}}+\frac{1}{6}(\delta_{-}p_{n,t_{2m-1}}+3\delta_{-}(u_nv_{n,t_{2m-1}}))-\frac{1}{3}\delta_{-}w_{n,t_3,t_{2m-1}}=0,\label{2}
\end{eqnarray}
Considering (\ref{d1}), from (\ref{1}) we have
\begin{eqnarray*}
4\delta_{-}v_{n,t_3}+\delta_{-}q_n+\frac{6}{h}(\delta_{-}u_n)(\delta_{-}v_n)-\frac{3}{2}(\delta_{-}u_n)(\delta_{-}v_n)^2=0.
\end{eqnarray*}
Integrate above equation with $x$, we get
\begin{eqnarray}
4\delta_{-}w_{n,t_3}+\delta_{-}p_n+\frac{3}{h}(\delta_{-}v_n)^2-\frac{1}{2}(\delta_{-}v_n)^3=0.\label{3}
\end{eqnarray}
Eliminating $t_3$ from (\ref{2}) and (\ref{3})
\begin{eqnarray*}
&& 4\delta_{-}v_{n,t_{2m+1}}+\delta_{-}p_{n,t_{2m-1}}+2\delta_{-}(u_nv_{n,t_{2m-1}})+(\delta_{+}u_{n})(\delta_{-}v_{n,t_{2m-1}})=0,
\end{eqnarray*}
or
\begin{eqnarray*}
&& 4v_{n,t_{2m+1}}+p_{n,t_{2m-1}}+2u_nv_{n,t_{2m-1}}+\delta_{-}^{-1}((\delta_{+}u_{n})(\delta_{-}v_{n,t_{2m-1}}))=0.
\end{eqnarray*}
Thus we get a new system
\begin{eqnarray}
&& \delta_{+}p_n=\frac{2}{h}\delta_{-}u_n-(\delta_{-}u_n)(\delta_{-}v_n),\label{eq1}\\
&& \delta_{+}q_n=\frac{2}{h}\delta_{-}p_n-(\delta_{-}u_n)^2-(\delta_{-}p_n)(\delta_{-}v_n),\label{eq2}\\
&& \delta_{+}r_n=\frac{2}{h}\delta_{-}q_n-3(\delta_{-}p_n)(\delta_{-}u_n)-(\delta_{-}q_n)(\delta_{-}v_n),\label{eq3}\\
&& 4v_{n,t_3}+q_n+3u_n^2=0,\label{eq4}\\
&& 4u_{n,t_3}+r_n+6p_nu_n=0,\label{eq5}\\
&& 4v_{n,t_{2m+1}}+p_{n,t_{2m-1}}+2u_nv_{n,t_{2m-1}}+\delta_{-}^{-1}((\delta_{+}u_{n})(\delta_{-}v_{n,t_{2m-1}}))=0,\label{eq6}\\
&& 4u_{n,t_{2m+1}}+q_{n,t_{2m-1}}+2p_nv_{n,t_{2m-1}}+4u_nu_{n,t_{2m-1}}=0,\label{eq7}
\end{eqnarray}
If $m=1$, considering $t_1=x$ and $u_n=v_{n,x}$, $p_n=u_{n,x}$, $q_n=p_{n,x}$, $r_n=q_{n,x}$, (\ref{eq6})-(\ref{eq7}) are just (\ref{eq4})-(\ref{eq5}) and all these equations make an integrable discretization of the KdV equation (\ref{kdv1}).

Deriving (\ref{eq1}), (\ref{eq2}) with $t_{2m-1}$ once, we have
\begin{eqnarray}
 \quad p_{n,t_{2m-1}}&=&\delta_{+}^{-1}\big((\frac{2}{h}-v_{n+1}+v_n)\delta_{-}u_{n,t_{2m-1}}\big)-\delta_{+}^{-1}\big((u_{n+1}-u_n)(\delta_{-}v_{n,t_{2m-1}})\big)\nonumber\\
 &=&[\delta_{+}^{-1}(\frac{2}{h}-v_{n+1}+v_n)](\delta_{-}u_{n,t_{2m-1}})-[\delta_{+}^{-1}(u_{n+1}-u_n)](\delta_{-}v_{n,t_{2m-1}}),\\
 \quad q_{n,t_{2m-1}}&=&\delta_{+}^{-1}\big((\frac{2}{h}-v_{n+1}+v_n)\delta_{-}p_{n,t_{2m-1}}\big)-\delta_{+}^{-1}\big((p_{n+1}-p_n)(\delta_{-}v_{n,t_{2m-1}})\big)\nonumber\\
 &&-2\delta_{+}^{-1}\big((u_{n+1}-u_n)(\delta_{-}u_{n,t_{2m-1}})\big),\nonumber\\
 &=&[\delta_{+}^{-1}(\frac{2}{h}-v_{n+1}+v_n)\delta_{-}\delta_{+}^{-1}(\frac{2}{h}-v_{n+1}+v_n)-2\delta_{+}^{-1}(u_{n+1}-u_n)](\delta_{-}u_{n,t_{2m-1}})\nonumber\\
 &&-[\delta_{+}^{-1}\big(\delta_{-}\delta_{+}^{-1}(\frac{2}{h}-v_{n+1}+v_n)(u_{n+1}-u_n)\big)\nonumber\\
 &&+\delta_{+}^{-1}(\frac{2}{h}-v_{n+1}+v_n)\delta_{-}\delta_{+}^{-1}(u_{n+1}-u_n)](\delta_{-}v_{n,t_{2m-1}})
\end{eqnarray}
Here $[\delta_{+}^{-1}(\frac{2}{h}-v_{n+1}+v_n)]$ means an operator instead of $(\frac{2}{h}-v_{n+1}+v_n)$ being effected by $\delta_{+}^{-1}$,
Thus we get the following recursion system
\begin{equation}
4\left( \begin{array}{c}
v_n\\
u_n
\end{array} \right)_{t_{2m+1}}+\left(
\begin{array}{cc}
A & B  \\
C & D
\end{array}\right)
\left( \begin{array}{c}
v_n\\
u_n
\end{array} \right)_{t_{2m-1}}=0,\label{discrete operator}
\end{equation}
with initial system
\begin{equation}
\left( \begin{array}{c}
v_n\\
u_n
\end{array} \right)_{t_{1}}=
\left( \begin{array}{c}
u_n\\
\delta_{+}^{-1}(\frac{2}{h}(u_{n+1}-u_n)-(u_{n+1}-u_n)(v_{n+1}-v_n))
\end{array} \right)
\end{equation}
where
\begin{eqnarray}
 \quad A&=&[2u_n+\delta_{-}^{-1}(u_{n+1}+u_n)\delta_{-}-\delta_{+}^{-1}(u_{n+1}-u_n)\delta_{-}]\\
 \quad B&=&[\delta_{+}^{-1}(\frac{2}{h}-v_{n+1}+v_n)\delta_{-}]\\
 \quad C&=&[2\big(\delta_{+}^{-1}(\frac{2}{h}-v_{n+1}+v_n)(u_{n+1}-u_n)\big)-\delta_{+}^{-1}\big(\delta_{-}\delta_{+}^{-1}(\frac{2}{h}-v_{n+1}+v_n)(u_{n+1}-u_n)\big)\delta_{-}\nonumber\\
 \quad &&-\delta_{+}^{-1}(\frac{2}{h}-v_{n+1}+v_n)\delta_{-}\delta_{+}^{-1}(u_{n+1}-u_n)\delta_{-}\\
 \quad D&=&[4u_n+\delta_{+}^{-1}(\frac{2}{h}-v_{n+1}+v_n)\delta_{-}\delta_{+}^{-1}(\frac{2}{h}-v_{n+1}+v_n)\delta_{-}-2\delta_{+}^{-1}(u_{n+1}-u_n)\delta_{-}]
\end{eqnarray}

As $h\rightarrow 0$, $\delta_{+}\rightarrow 2 E^0$($E^0$ is the identity operator), $\frac{1}{h}\delta_{-}\rightarrow \partial_x$ and $\delta_{+}^{-1}\rightarrow \frac{1}{2} E^0$ , $h\delta_{-}^{-1}\rightarrow \partial_x^{-1}$, thus

\begin{eqnarray}
A&\rightarrow&[2u+2\partial_x^{-1}u\partial_x]\\
B&\rightarrow&[\partial_x]\\
C&\rightarrow&[2u_x]\\
D&\rightarrow&[4u+\partial_x^2]
\end{eqnarray}
Then (\ref{discrete operator}) become
\begin{equation}
4\left( \begin{array}{c}
v\\
u
\end{array} \right)_{t_{2m+1}}+\left(
\begin{array}{cc}
2u+2\partial_x^{-1}u\partial_x & \partial_x  \\
2u_x & 4u+\partial_x^2
\end{array}\right)
\left( \begin{array}{c}
v\\
u
\end{array} \right)_{t_{2m-1}}=0,\label{smooth operator2}
\end{equation}
Noting that $u=v_x$, (\ref{smooth operator2}) is equivalent to (\ref{smooth operator1}). Thus, for the KdV hierarchy, we have the following proposition.
\begin{prop}
The differential-difference system (\ref{discrete operator}) is an intgerable discretization of the KdV hierarchy (\ref{kdv1})-(\ref{smooth operator1}). 
As $h\rightarrow 0$, the discrete  operator in (\ref{discrete operator}) converges to the recursion operator in (\ref{smooth operator1}).
\end{prop}
It is known that the discrete KdV equation can be understood as a generalized Volterra system \cite{sur,akns8}. 
However, the recursion operator derived in this paper deviates from the known recursion operators of the Volterra hierarchy \cite{sur,akns8,akns15}. 
Instead, the discrete system bears a closer resemblance to the one presented in \cite{schiff}, where a full-discretized KdV hierarchy is derived through the loop group approach.

\section{Discussions and Conclusion}\label{sec4}
In this paper, we have showcased the derivation of unified bilinear forms for both the AKNS hierarchy and the KdV hierarchy originating from their recursive expressions. Furthermore, utilizing these bilinear forms, we were able to deduce their corresponding bilinear B\"acklund transformations.
Building upon the framework of these B\"acklund transformations, we proceeded to derive integrable discretizations for the aforementioned hierarchies. This process led us to the acquisition of their discrete recursion operators. Through this comprehensive approach, we have effectively demonstrated a systematic method for discretizing recursion operators by capitalizing on the inherent equivalence between recursion operators and unified bilinear forms. To encapsulate our methodology, we present the following diagram
\setlength{\unitlength}{1mm}
\begin{center}
\begin{picture}(100,100)\thicklines
\put(10,65){
\newsavebox{\processa}
\savebox{\processa}(30,20)
{\put(0,20){\line(1,0){30}}
\put(0,0){\line(0,1){20}}
\put(0,0){\line(1,0){30}}
\put(30,0){\line(0,1){20}}
\put(7,11){Recursion}
\put(8,6){operator}}
\usebox{\processa}}

\put(45,75){\vector(1,0){17}}

\put(30,62){\line(0,-1){1}}
\put(30,60){\line(0,-1){1}}
\put(30,58){\line(0,-1){1}}
\put(30,56){\line(0,-1){1}}
\put(30,54){\line(0,-1){1}}
\put(30,52){\line(0,-1){1}}
\put(30,50){\line(0,-1){1}}
\put(30,48){\line(0,-1){1}}
\put(30,46){\line(0,-1){1}}
\put(30,44){\line(0,-1){1}}
\put(30,42){\line(0,-1){1}}
\put(30,40){\line(0,-1){1}}
\put(30,38){\vector(0,-1){1}}

\put(10,15){
\newsavebox{\processb}
\savebox{\processb}(30,20)
{\put(0,20){\line(1,0){30}}
\put(0,0){\line(0,1){20}}
\put(0,0){\line(1,0){30}}
\put(30,0){\line(0,1){20}}
\put(9,13){Discrete}
\put(8,8){recursion}
\put(9,3){operator}}
\usebox{\processb}}

\put(60,65){
\newsavebox{\processc}
\savebox{\processc}(30,20)
{\put(0,20){\line(1,0){30}}
\put(0,0){\line(0,1){20}}
\put(0,0){\line(1,0){30}}
\put(30,0){\line(0,1){20}}
\put(9,11){Unified}
\put(5,6){bilinear form}}
\usebox{\processc}}

\put(62,25){\vector(-1,0){17}}
\put(82,50){integrable}
\put(82,45){discretization}
\put(78,63){\vector(0,-1){26}}

\put(60,15){
\newsavebox{\processd}
\savebox{\processd}(30,20)
{\put(0,20){\line(1,0){30}}
\put(0,0){\line(0,1){20}}
\put(0,0){\line(1,0){30}}
\put(30,0){\line(0,1){20}}
\put(9,13){Unified}
\put(5,8){bilinear form}
\put(3,3){in discrete case}}
\usebox{\processd}}

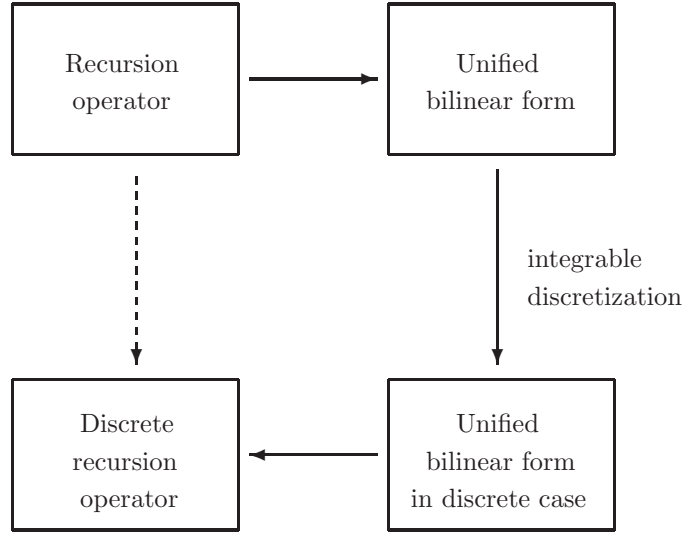
\put(11,1){Fig 1. Procedure of discretizing recursion operators}

\end{picture}
\end{center}

There is still further work to be undertaken in this direction. One important aspect is to verify whether the obtained recursion operators indeed serve as strong symmetry operators for the integrable discretizations of the AKNS and KdV hierarchies. This validation process is crucial to establish the effectiveness and compatibility of the derived discretizations. 

As a closing statement, it is worth noting a relevant precedent in \cite{Hu3}, where Springael, Hu, and Loris successfully derived a recursion operator for the Ito hierarchy from its unified bilinear form. Furthermore, in \cite{liu}, Liu was able to demonstrate that this recursion operator does indeed function as a strong symmetry operator. These examples underscore the significance of establishing the connection between recursion operators and strong symmetry operators in integrable discretizations.

In conclusion, while this study has provided valuable insights into the discretization of recursion operators through the equivalence of recursion operators and unified bilinear forms, further investigations are essential to solidify the validity and implications of these findings within the broader context of integrable systems. Moreover, we plan to apply this procedure to other integrable systems, including the Boussinesq hierarchy, the Sawada-Koterra hierarchy, and the Ramani hierarchy.

\subsection*{Acknowledgements}
This work was supported by the National Natural Science Foundation of China (Grant
no. 11931017, 12071447, 12175155, 12371251) and the Natural Science Foundation of Jiangsu Province (Grant no. BK20211266).

\label{lastpage}

\begin{thebibliography}{99}
\bibitem{hirota71}Hirota R,  Exact Solution of the Korteweg-de Vries Equation for Multiple Collisions of Solitons,
{\it Phys.  Rev. Lett.} {\bf 27}, 1192-1994, 1971.
\bibitem{hirota721}Hirota R, Exact Solution of the modified Korteweg-de Vries Equation
for Multiple Collisions of Solitons, {\it J.  Phys.  Soc.  Jpn.} {\bf 33}, 1456-1458, 1972.
\bibitem{hirota722} Hirota R, Exact Solution of the Sine-Gordon Equation for
Multiple Collisions of Solitons, {\it J. Phys.  Soc.  Jpn.} {\bf 33}, 1459-1463, 1972.
\bibitem{hirota73} Hirota R, Exact envelope-soliton solutions of a nonlinear wave
equation, {\it J.  Math.  Phys.}  {\bf 14}, 805-809, 1973.
\bibitem{Newell} Newell A C, Solitons in Mathematics and Physics, SIAM Philadelphia, 1985.
\bibitem{SA}Sato M and Sato Y, Soliton equations as dynamical systems on infinite dimensional Grassmann manifold, Nolinear Partial Differential Equations in Applied Sciences, Lax P D, Fujita H  and Strang G, {\it Math. Stud.}, 259-271, 1982.
\bibitem{DKJM} Date E, Kashiwara M, Jimbo M and  Miwa T, Nonlinear integrable systems-classical theroy and quantumn theory, Proc. RIMS Symp. Miwa T and Jimbo M (eds.) Singapore: World Scientific, 39-119, 1983.
\bibitem{JM} Date E and Jimbo M, Solitons and infinite dimensional
Lie algebras, {\it Publ.  RIMS. Kyoto Univ. } {\bf 19} 943-1001, 1983.
\bibitem{Ma1} Matsuno Y, Bilinearization of nonlinear evolution equations,
{\it J.  Phys.  Soc.  Jpn.} {\bf 48}, 2138-2143, 1980.
\bibitem{Ma2} Matsuno Y, Bilinearization of nonlinear evolution equations II. Higher-order
 modified Korteweg-de Vries equations,
{\it J.  Phys.  Soc.  Jpn.} {\bf 49}, 787-794, 1980.
\bibitem{Ma3} Matsuno Y, Bilinearization of nonlinear evolution equations IV. Higher-order
Benjamin-Ono equations, {\it J.  Phys.  Soc.  Jpn.} {\bf 49}, 1584-1592, 1980.
\bibitem{Ma4} Matsuno Y, Bilinear Transformation Method. New York: Academic,1984.
\bibitem{Hu1} Hu X B and Li Y, Bilinearization of KdV, MKdV and classical Boussinesq hierarchies:Report on International Conference on Nonlinear Physics (Shanghai 24-30 April 1989).
\bibitem{Hu2} Liu Q M, Hu X B and Li Y, Rational solutions of classical Boussinesq hierarchy, {\it J. Phs. A:Math. Gen.} {\bf 23},585-591, 1990.
\bibitem{olver}Olver P J, Evolution equations possessing infinitely many symmetries, {\it J.  Math.  Phys.} {\bf 18}, 1212, 1977.
\bibitem{zk} Zakharov V E and Konopelchenko B G, On the theory of recursion operator, {\it Commun. Math. Phys.} {\bf 94(4)}, 483-509, 1984.
\bibitem{Fuchssteiner}Fuchssteiner B, Application of hereditary symmetries to nonlinear evolution equations, {\it Nonlinear
Anal. Theory. Meth. Appl.} {\bf 3}, 849-862, 1979.
\bibitem{sf}Santini P M and Fokas A S, Recursion operators and bi-Hamiltonian structures in multidimensions I, {\it Commun. Math. Phys.} {\bf 115}, 375-419, 1988.
\bibitem{fs}Fokas A S and  Santini P M, Recursion operators and biHamiltonian structures in multidimensions II, {\it Commun. Math. Phys.} {\bf 116}, 449-474, 1988.
\bibitem{akns} Ablowitz M J, Kaup D J, Newell A C and Segur H, The inverse scattering transform-Fourier analysis for nonlinear problems, {\it Stud. Appl. Math.} {\bf 53}, 249-315, 1974.
\bibitem{fn} Flaschka H and Newell A, Integrable systems of nonlinear evolution equations,  {\it Dynamical Systems, Theory and Applications}, 355-440, 2005.
\bibitem{ff}Fuchssteiner B and Fokas A S, Symplectic structures, their B\"aklund transformations and hereditary symmetries, {\it Physica D} {\bf 4(1)}, 47-66, 1981.
\bibitem{fa}Fokas A S and Anderson R L, On the use of isospectral eigenvalue problems for obtaining hereditary symmetries for Hamiltonian systems. {\it J. Math. Phys.} {\bf 23}, 1066, 1982.
\bibitem{fg}Fokas A S and Gelfand I M, Bi-Hamiltonian structures and integrability: Important developments in soliton theory, Springer Berlin Heidelberg, 259-282, 1993.
\bibitem{oevel}Oevel W and Popowicz Z, The bi-Hamiltonian structure of fully supersymmetric Korteweg-de Vries systems. {\it Commun. Math. Phys.} {\bf 139}, 441-460, 1991.
\bibitem{wang1}Sanders J A and Wang J P, On recursion operators, {\it Physica D} {\bf 149}, 1-10, 2001.
\bibitem{wang2}Mikhailov A V, Wang J P and Xenitidis P, Recursion operators, conservation laws, and integrability conditions for difference equations, {\it Theor. Math. Phys.} {\bf 167}, 421-443, 2011.
\bibitem{wang2019}Carpentier S, Mikhailov A V and  Wang J P, Rational Recursion Operators for Integrable Differential–Difference Equations, {\it Commun. Math. Phys.} {\bf 370}, 807–851, 2019.
\bibitem{Hu3} Springael J, Hu X B and Loris I,  Bilinear characterization of higher order Ito-equations, {\it J. Phys. Soc. Jpn.} {\bf 65}, 1222-1226, 1996.
\bibitem{Hu4} Hu X B and Bullough R, A B\"aklund transformation and nonlinear superposition formula of the Caudrey-Dodd-Gibbon-Kotera-Sawada hierarchy, {\it J. Phys. Soc. Jpn.} {\bf 67}, 772-777, 1998.
\bibitem{AL} Ablowitz M J and Ladik J F, Nonlinear differential-difference equations, {\it J. Math. Phys.} {bf 16}, 598-605, 1975.
\bibitem{hr1} Hirota R, Nonlinear partial difference equations I. A difference analogue of the Korteweg-de Vries equation, {\it J. Phys. Soc. Jpn.} {\bf 43}, 1424-1433, 1977.
\bibitem{hr2} Hirota R, Nonlinear partial difference equations II. Discrete-time Toda equation, {\it J. Phys. Soc. Jpn.} {\bf 43}, 2074-2078, 1977.
\bibitem{hr3} Hirota R, Nonlinear partial difference equations III. Discrete Sine-Gordon equation, {\it J. Phys. Soc. Jpn.} {\bf 43}, 2079-2086, 1977.
\bibitem{ncwq} Nijhoff F W, Capel H W, Wiersma G L and  Quispel G R W, Linearizing integral transform and partial difference equations, {\it Phys. Lett. A} {\bf 103}, 293-297, 1984.
\bibitem{qncl} Quispel G R W, Nijhoff F W, Capel H W and Van-der Linden J, Linear integral equations and nonlinear differrencedifference
equations, {\it Physica A} {\bf 125}, 344-380, 1984.
\bibitem{levi1} Levi D and Benguria R, B\"acklund transformations and nonlinear differential difference equations, {\it P. Natl. Acad. Sci.(USA)} {\bf 77}, 5025-5027, 1980.
\bibitem{levi2} Levi D, Nonlinear differential difference equations as B\"acklund transformations, {\it J. Phys. A: Math. Gen.} {\bf 14}, 1083-1098, 1981.
\bibitem{sur} Suris Y B, The problem of integrable discretization: Hamiltonian approach, Birkh\"auser Basel, 2003.
\bibitem{schiff}Schiff J, Loop groups and discrete KdV equations, {\it Nonlinearity} {\bf 16}, 257, 2002.
\bibitem{bs}Bobenko A I and Suris Y B, Discrete differential geometry:integrable structure.American Mathematical Society, 2008.
\bibitem{ablowitz2004}Ablowitz M J, Prinari B and Trubatch A D, Discrete and continuous nonlinear Schrödinger systems, Cambridge University Press, 2004.
\bibitem{Hietarinta2016}Hietarinta J, Joshi N and Nijhoff F W, Discrete systems and integrability, Cambridge university press, 2016.
\bibitem{levi2023}Levi D, Winternitz P and Yamilov R I, Continuous Symmetries and Integrability of Discrete Equations, American Mathematical Society, Centre de Recherches Mathématiques, 2023.
\bibitem{feng1}Feng B F, Maruno K and Ohta Y, Integrable semi-discretizations of the reduced Ostrovsky equation, {\it J. Phys. A: Math. Theo.} {\bf 48}, 135203, 2015.
\bibitem{feng2}Chen J, Chen Y, Feng B F, Maruno K and Ohta Y, An integrable semi-discretization of the coupled Yajima-Oikawa system, {\it J. Phys. A: Math. Theo.} {\bf 49}, 165201, 2016.
\bibitem{suris1}Petrera M and Suris Y B, On the Hamiltonian structure of Hirota-Kimura discretization of the Euler top, {\it Math. Nachr.} {\bf 283}, 1654-1663, 2010.
\bibitem{Tsuchida}Tsuchida T and Dimakis A, On a (2+ 1)-dimensional generalization of the Ablowitz-Ladik lattice and a discrete Davey-Stewartson system, {\it J. Phys. A: Math. Theo.} {\bf 44}, 325206, 2011.
\bibitem{veni} Veni S S and Latha M M, A generalized Davydov model with interspine coupling and its integrable discretization, {\it Physica
Scripta} {\bf 86}, 025003, 2012,.
\bibitem{yu1}Vinet L, Yu G F and Zhang Y N, On an integrable system related to the relativistic Toda lattice-B\"cklund transformation and integrable discretization, {\it J. Differ. Equ. Appl.} {\bf 21}, 403-417, 2015.
\bibitem{yu2}Yu G F and Xu Z W, Dynamics of a differential-difference integrable (2+1)-dimensional system, {\it Phys. Rev. E} {\bf 91(6)}, 062902, 2015.
\bibitem{zhang1} Zhang Y, Tam H W and Hu X B, Integrable discretization of time and its application on the Fourier pseudospectral
method to the Kortewegde Vries equation, {\it J. Phys. A: Math. Gen.} {\bf 47}, 045202, 2014.
\bibitem{zhang2} Zhang Y, Hu X B and Tam H W, Integrable discretization of nonlinear Schr\"odinger equation and its application with
Fourier pseudo-spectral method,  {\it Numer. Algorithms} {\bf 69}, 839-862, 2015.
\bibitem{zhang3}Zhang Y N, Chang X K, Hu J, Hu X B and Tam H W, Integrable discretization of soliton equations via bilinear method and B\"acklund transformation, {\it Sci. China Math.} {\bf 58}, 279-296, 2015.
\bibitem{zhang4}Zhang Y and Tian L, An integrable semi-discretization of the Boussinesq equation, {\it Phys. Lett. A} {\bf 380(43)}, 3575-3582, 2016.
\bibitem{liu1}Huang W H, Xue L L and Liu Q P, Integrable discretizations for classical Boussinesq system, {\it J. Phys. A: Math. Theor.} {\bf 54(4)}, 045201, 2021.
\bibitem{zhu1}Zhao H Q,  Yuan J Y and Zhu Z N, Integrable semi-discrete Kundu-Eckhaus equation: Darboux transformation, breather, rogue wave and continuous limit theory, {\it J. Nonlinear Sci.} {\bf 28}, 43-68, 2018.
\bibitem{Ablow2} Ablowitz M J and  Segur H, Solitons and the Inverse Scattering Transform, SIAM Philadelphia, 1985.
\bibitem{calogero}Calogero F and Degasperis A,  Nonlinear evolution equations solvable by the inverse spectral transform I, {\it Nuovo. Cimento. B} {\bf 32} 201-242, 1976.
\bibitem{AL1} Ablowitz M J and Ladik J F, Nonlinear differential-difference equations and Fourier analysis, {\it J. Math. Phys.} {\bf 16}, 1011-1018, 1976.

\bibitem{akns51} Van Moerbeke P and Mumford D, The spectrum of difference operators and algebraic curves, {\it Acta Mathematica} {\bf 143(1)}, 93-154,  1979.
\bibitem{akns52} Moser J, Three integrable Hamiltonian systems connected with isospectral deformations, {\it Adv. Math.} {\bf 16(2)}, 197-220, 1975.
\bibitem{akns8} Kako F and Mugibayashi N, Complete integrability of general nonlinear differential-difference equations solvable by the inverse method I, {\it Prog. Theor. Phys.} {\bf 60(4)}, 975-984, 1978.
\bibitem{akns9} Kako F and Mugibayashi N, Complete integrability of general nonlinear differential-difference equations solvable by the inverse method II, {\it Prog. Theor. Phys.} {\bf 61(3)}, 776-790, 1979.
\bibitem{akns11} Bruschi M, Manakov S V, Ragnisco O and Levi D, The nonabelian Toda lattice: discrete analogue of the matrix Schr\"odinger spectral problem, {\it J. Math. Phys.} {\bf 21(12)}, 2749-2753, 1980.
\bibitem{zhangdajun1}Zhang D J and Chen S T, Symmetries for the Ablowitz-Ladik Hierarchy: Part II. Integrable Discrete Nonlinear Schr\"odinger Equations and Discrete AKNS Hierarchy, {\it Stud.  Appl. Math.} {\bf 125(4)}, 419-443, 2010.
\bibitem{akns12} Konopelchenko B G,  Elementary B\"acklund transformations, nonlinear superposition principle and solutions of the integrable equations, {\it Phys. Lett. A} {\bf 87}, 445-448, 1982.
\bibitem{akns13} Chudnovsky D V and Chudnovsky G V,  B\"acklund transformation as a method of decomposition and reproduction of two-dimensional nonlinear systems, {\it Phys. Lett. A} {\bf 87}, 325-329, 1982.
\bibitem{akns14} Gerdzhikov V S and Ivanov M I, Hamiltonian structure of multicomponent nonlinear Schr\"odinger equations in difference form, {\it Theor. Math. Phys.} {\bf 52}, 676-685, 1982.
\bibitem{akns15} Zhang H, Tu G  Z, Oevel W and Fuchssteiner B, Symmetries, conserved quantities, and hierarchies for some lattice systems with soliton structure, {\it J. Math. Phys.} {\bf 32}, 1908-1918, 1991.
\bibitem{akns16} Merola I, Ragnisco O  and Tu G Z,  A novel hierarchy of integrable lattices, {\it Inverse. Probl.} {\bf 10}, 1315-1334, 1994.

\bibitem{nijhoff}Nijhoff F and Capel H, The discrete Korteweg-de Vries equation, {\it Acta Appl. Math.} {\bf 39}, 133-158, 1995.
\bibitem{ohta}Ohta Y and Hirota R, A discrete KdV equation and its Casorati determinant solution,  {\it J. Phys. Soc. Jpn.} {\bf 60}, 2095-2095, 1991.
\bibitem{liu}Liu Q P, Hamiltonian structures for Ito's equation, {\it Phys. Lett. A} {\bf 277(1)}, 31-34, 2000.

\end{thebibliography}
\end{document}